\documentclass[aps,prl,twocolumn,groupedaddress,showpacs]{revtex4}
\usepackage{amsmath,amsfonts,eucal}

\renewcommand{\vec}[1]{\boldsymbol{#1}}

\makeatletter

{\catcode`\|=\active
  \gdef\Braket#1{\left<\mathcode`\|"8000\let|\BraVert {#1}\right>}}
\def\BraVert{\egroup\,\mid@vertical\,\bgroup}
{\catcode`\|=\active
  \gdef\set#1{\mathinner{\lbrace\,{\mathcode`\|"8000\let|\midvert #1}\,\rbrace}}
  \gdef\Set#1{\left\{\:{\mathcode`\|"8000\let|\SetVert #1}\:\right\}}}
\def\midvert{\egroup\mid\bgroup}
\def\SetVert{\egroup\;\mid@vertical\;\bgroup}
\begingroup
 \edef\@tempa{\meaning\middle}
 \edef\@tempb{\string\middle}
\expandafter \endgroup \ifx\@tempa\@tempb

\def\mid@vertical{\middle|}
\else
 \let\mid@vertical\vrule
\fi
\makeatother

\begin{document}

\title{Slowing the Decay of an Excited Atom by the Presence of a Detector}
\author{D. R. Fredkin}
\affiliation{Department of Physics, University of California, San Diego, La Jolla, CA
92093}
\author{Amiram Ron}
\affiliation{Department of Physics, Technion---Israel Institute of Technology, Haifa,
ISRAEL 32000}
\date{\today}

\begin{abstract}

The decay of an excited is shown to slow down in the presence of a photo
detector. This is similar to the behavior of an atom in a mistuned cavity,
and under the conditions of the quantum Zeno effect. No external perturbing
field is applied to measure the survival probability of the unstable state.
\end{abstract}

\pacs{03.65.Xp, 03.65.Yz, 06.20.Dk}
\maketitle












\section{Introduction}

It has long been known that the rate of decay of an excited atom can be
decreased, or increased, relative to its vacuum value, by changing the \emph{%
environment} of the atom, e.g., placing it between two mirrors\cite%
{Eschner2001}. The rate of decay can also be changed by \emph{measuring} the
state of the system sufficiently frequently\cite{Koshino2005}, the so called
quantum Zeno effect. In the present paper we will show that, furthermore, an
excited atom may be prevented from returning to its ground state merely by
being \emph{watched} by a appropriate detector.

As early as 1946 Purcell\cite{Purcell1946} made the observation that the
spontaneous emission rate of an atom is increased when placed in a cavity
tuned to its transition frequency. It was later argued\cite{Kleppner1971}
that this rate can be decreased when the cavity is mistuned, or in principle
even may completely inhibited\cite{Kleppner1981}. This effect was first
observed\cite{Drexhage1974} for dye molecules near a conducting plane, and
later on for Rydberg atoms\cite{Hulet1985}, and in the visible range\cite%
{Heinzen1987}.

In a theoretical paper, Misra and Sudarshan\cite{Misra1977} pointed out
that, assuming that measurements are described by von Neumann's \emph{state
reduction postulate,} frequently measuring an unstable state of a system
inhibits its decay. It was first demonstrated experimentally by Itano et al
\cite{Itano1990} that a quantum transition of an atomic system is restrained
by frequently measuring the state of the atom. Their analysis of the
experiments in terms of the \emph{collapse of the wave function} following
the measurements agrees remarkably well with the experimental results. It
was later shown \cite{Schenzle1991} that analysis of this experiment does
not require the projection postulate; standard quantum dynamical analysis
can account for the time evolution of the atom under the influence of the
external field, which ``measures'' the state of the system. A decade later
Fischer et al\cite{Fischer2001} observed that the escape rate of trapped
cold atoms by tunneling is either decelerated or accelerated by frequently
measuring the number of atoms which remained trapped. The quantum Zeno
effect was understood as due to an \emph{active} intervention of a measuring
entity in the \emph{early} evolution of the system; the survival probability
of the quantum state is $1-at^{2}$, where $a$ is a constant, for
sufficiently short time, $t$ (see e.g. \cite%
{Khalfin1958,Peres1980b,Facchi2001} for a detailed analysis).

In the present paper we investigate the time evolution of an excited atom in
the presence of a detector, without active intervention of an external field
with the atom, and without restricting ourselves to the early time dynamics.
We demonstrate that the mere presence of a detector slows down the
transition of the atom to the ground state. We start with a model
Hamiltonian of the system composed of an atom, the radiation field, and a
detector made of one or many atoms, which can be ionized by the radiation
emitted by the excited atom. The radiation field simply mediates between the
excited atom and the atoms of the detector, which are initially in their
ground state. We investigate the evolution of the entire system,
generalizing the standard procedure of Weisskopf and Wigner\cite{WW1930}.


\section{Model Hamiltonian}

We consider a two level atom at the origin of the coordinate system, and a
collection of detector atoms, perhaps constituting a spherical shell
detector near radius $R$. The Hamiltonian of the atom, with ground state $%
\mathinner{|{l}\rangle}$, of energy $\epsilon_{l}=0$, and the excited state $%
\mathinner{|{u}\rangle}$, of energy $\epsilon_{u}=\hbar \omega_{0}$, is
expressed as $H_{a}=\hbar \omega _{0} \mathinner{|{u}\rangle} %
\mathinner{\langle{u}|}$. The atom is coupled to the electromagnetic field,
whose normal modes, described in terms of the creation and annihilation
operators $b_{\vec{k},\lambda}^{\dagger}$, and $b_{\vec{k},\lambda }$, have
wave vector $\vec{k}$, polarization unit vector $\widehat{\boldsymbol{%
\epsilon}}_{\vec{k},\lambda }$, with $\vec{k}\cdot \widehat{\boldsymbol{%
\epsilon }}_{\vec{k},\lambda }=0$, and frequencies $\omega _{k}=ck$. The
Hamiltonian of the field is $H_{w}=\hbar \sum_{\vec{k},\lambda }\omega
_{k}b_{\vec{k},\lambda }^{\dagger }b_{\vec{k},\lambda }$, and its coupling
to the atom is given by $H_{wa}=-\vec{d}_{a}\cdot \vec{E}\left( \vec{r}%
,t\right) $. Here $\vec{E}\left( \vec{r},t\right) $ is the radiation
electric field, and $\vec{d}_{a}=\widehat{\boldsymbol{p}}_{a}d_{A}$ is the
atom's dipole moment. In the rotating wave approximation the coupling
Hamiltonian is
\begin{equation}
H_{wa}=\hbar \sum_{\vec{k},\lambda }\left( \alpha _{\vec{k},\lambda }%
\mathinner{|{u}\rangle} \mathinner{\langle{l}|} b_{\vec{k,}\lambda }+\alpha
_{\vec{k},\lambda }^{\ast }b_{\vec{k},\lambda }^{\dagger }%
\mathinner{|{l}\rangle} \mathinner{\langle{u}|}\right) ,  \label{Hcw-1}
\end{equation}
where
\begin{equation}
\alpha _{\vec{k},\lambda }=-i\mu _{a}\sqrt{2\pi \omega _{k}/\hbar V}\left(
\widehat{\boldsymbol{p}}_{a}\cdot\widehat{\boldsymbol{\epsilon}}_{\vec{k}%
,\lambda }\right) ,  \label{alp-1}
\end{equation}
$V$ is the quantization volume, and $\mu _{a}=%
\mathinner{\langle{u | d_{A} |
l}\rangle}$. The detector is made of identical atoms, located at positions $%
\vec{r}_{i}$, whose ionization energy $E_{I}=\hbar \omega _{I}$, is much
smaller than $\hbar \omega _{0}$. The ground state of a detector atom is $%
\mathinner{|{g}\rangle}$, with energy $\epsilon_{g}=0$, the ionized states
are $\mathinner{|{c}\rangle}$, with energies $\epsilon _{c}=\hbar \omega _{c}
$, and the detector's Hamiltonian is $H_{d}=\hbar \sum_{i,c}\omega _{c}%
\mathinner{|{c}\rangle} _{i}\mathinner{\langle{c}|}_{i}$. The dipole
coupling of the $i^\text{th}$ detector atom to the field is $H_{wd,i}=-\vec{d%
}_{i}\cdot \vec{E}\left( \vec{r}_{i},t\right) $, where $\vec{d}_{i}=\widehat{%
\boldsymbol{p}}_{d}d_{D}$ is the dipole moment operator of the detector's
atom. The detector-field coupling Hamiltonian in rotating wave approximation
is
\begin{multline}
H_{wd} =\hbar \sum_{i,c}\sum_{\vec{k},\lambda }\Big\{ g_{\vec{k}%
,\lambda,c}e^{i\vec{k}\cdot\vec{r}_{i}}\mathinner{|{c}\rangle}_i%
\mathinner{\langle{g}|}_{i}b_{\vec{k},\lambda} \\
{}+g_{\vec{k},\lambda, c}^{\ast }e^{-i\vec{k}\cdot\vec{r}_{i}}b_{\vec{k}%
,\lambda}^{\dagger} \mathinner{|{g}\rangle}_{i}\mathinner{\langle{c}|}_{i}%
\Big\} ,  \label{Hd-2}
\end{multline}
where
\begin{equation}
g_{\vec{k,}\lambda ,c}=-i\mu _{c}\sqrt{2\pi \omega _{k}/\hbar V}\left(
\widehat{\boldsymbol{p}}_{d}\cdot\widehat{\boldsymbol{\epsilon }}_{\vec{k}%
,\lambda }\right) ,  \label{gkl-1}
\end{equation}
and $\mu _{c}=\mathinner{\langle{c | d_{D} |g}\rangle}$.

\section{Single atom detector}

We first consider a single atom detector at $\vec{r}$. The initial state of
the system is assumed to be: the atom is excited, while the field modes are
empty, and the detector is in its ground state. We are interested in the
evolution in time of the system from the initial state. The states available
to the system are

\begin{enumerate}
\item \renewcommand{\labelenumi}{(\roman{enumi})}

\item the initial state $\mathinner{|{\phi _{0}}\rangle} =%
\mathinner{|{u,\left\{
n_{\vec{k},\lambda}=0\right\} ,g}\rangle}$, of amplitude $A_{0}$,

\item single photon states
\begin{equation*}
\mathinner{|{\phi _{\vec{k},\lambda }}\rangle} =\mathinner{|{l;\left\{
n_{\vec{k},\lambda }=1,n_{\vec{k}^{\prime},\lambda^{\prime} \neq
\vec{k},\lambda }=0\right\} ;g}\rangle},
\end{equation*}
of amplitude $A_{\vec{k},\lambda }$, and

\item ionization states $\mathinner{|{\phi _{c}}\rangle} =%
\mathinner{|{ l;\left\{
n_{\vec{k},\lambda }=0\right\} ;c}\rangle}$, of amplitude $A_{c}$.
\end{enumerate}

Here $n_{\vec{k,}\lambda }$ is the number of photons in the mode.
Simplifying the notation by $\vec{k,}\lambda \rightarrow k$ and $\widetilde{g%
}_{kc}=g_{\vec{k},\lambda ,c}e^{i\vec{k}\cdot\vec{r}}$, the state of the
system is then
\begin{equation}
\mathinner{|{\Psi \left( t\right)}\rangle} =A_{0}(t)\mathinner{|{\phi
_{0}}\rangle} +\sum_{k}A_{k}(t)\mathinner{|{\phi _{k}}\rangle}
+\sum_{c}A_{c}(t)\mathinner{|{\phi _{c}}\rangle},  \label{ps-2}
\end{equation}
and the equations of motion for the amplitudes are:
\begin{gather}
\left( \frac{d}{dt}+i\omega _{0}\right) A_{0}\left( t\right)
=-i\sum_{k}\alpha _{k}A_{k}\left( t\right) ,  \label{ax-2} \\
\left( \frac{d}{dt}+i\omega _{k}\right) A_{k}\left( t\right) =-i\alpha
_{k}^* A_{0}\left( t\right) -i\sum_{c}\widetilde{g}_{kc}^{\ast }A_{c}\left(
t\right) ,  \label{b-1} \\
\left( \frac{d}{dt}+i\omega _{c}\right) A_{c}\left( t\right) =-i\sum_{k}%
\widetilde{g}_{kc}A_{k}\left( t\right) .  \label{ac-2}
\end{gather}

The Laplace transform of these equations, with $F\left( s\right)
=\int_{0}^{\infty }dt\,e^{-st}F\left( t\right) $, is
\begin{align}
\left( s+i\omega _{0}\right) A_{0}\left( s\right) =&1-i\sum_{k}\alpha
_{k}A_{k}\left( s\right) ,  \label{a-1} \\
\left( s+i\omega _{k}\right) A_{k}\left( s\right) =&-i\alpha _{k}^{\ast
}A_{0}\left( s\right) -i\sum_{c}\widetilde{g}_{kc}^{\ast }A_{c}\left(
s\right) ,  \label{a-2} \\
\left( s+i\omega _{c}\right) A_{c}\left( s\right) =&-i\sum_{k}\widetilde{g}%
_{kc}A_{k}\left( s\right) .  \label{a-3}
\end{align}
We use Eq.(\ref{a-2}) to eliminate the field amplitudes, i.e. we solve for $%
A_k(s)$:
\begin{multline}
A_{k}\left( s\right) = \\
-i\frac{\alpha _{k}^{\ast }}{\left( s+i\omega _{k}\right) }A_{0}\left(
s\right) -i\sum_{c^{\prime }}\frac{\widetilde{g}_{kc^{\prime }}^{\ast }}{%
\left( s+i\omega _{k}\right) }A_{c^{\prime }}\left( s\right) ,  \label{a}
\end{multline}
substitute it into Eq.(\ref{a-1}), and Eq.(\ref{a-3}) to obtain
\begin{equation}
\left( s+i\omega _{0}+K\left( s\right) \right) A_{0}\left( s\right)
=1-\sum_{c}M_{ac}\left( s\right) A_{c}\left( s\right)  \label{a-5}
\end{equation}
\begin{multline}
\left( s+i\omega _{c}\right) A_{c}\left( s\right) +\sum_{c^{\prime
}}N_{cc^{\prime }}\left( s\right) A_{c^{\prime }}\left( s\right) \\
=-M_{ca}\left( s\right) A_{0}\left( s\right) ,  \label{b}
\end{multline}
where the Laplace transforms of the propagators are
\begin{subequations}
\label{prop}
\begin{gather}
K\left( s\right) =\sum_{k}\frac{\left| \alpha _{k}\right| ^{2}}{\left(
s+i\omega _{k}\right) },  \label{ks-1} \\
M_{ac}\left( s\right) =\sum_{k}\frac{\alpha _{k}\widetilde{g}_{kc}^{\ast }}{%
\left( s+i\omega _{k}\right) },  \label{ks-2} \\
M_{ca}\left( s\right) =\sum_{k}\frac{\alpha _{k}^{\ast }\widetilde{g}_{kc}}{%
\left( s+i\omega _{k}\right) },  \label{ks-3} \\
N_{cc^{\prime }}\left( s\right) =\sum_{k}\frac{\widetilde{g}_{kc}\widetilde{g%
}_{kc^{\prime }}^{\ast }}{\left( s+i\omega _{k}\right) }.  \label{ks-4}
\end{gather}
Using Eq.(\ref{alp-1}), and Eq.(\ref{gkl-1}), and replacing
\end{subequations}
\begin{equation}
\sum_{k}\rightarrow \frac{V}{\left( 2\pi \right) ^{3}}\int
d^{3}k\,\sum_{\lambda },  \label{sd-5}
\end{equation}
Eq.(\ref{a-5}) and Eq.(\ref{b}) become
\begin{multline}
\left( s+i\omega _{0}+\left| \mu _{a}\right| ^{2}I\left( s\right) \right)
A_{0}\left( s\right) \\
=1-\mu _{a}J\left( s,\vec{r}\right) \sum_{c}\mu _{c}^{\ast }A_{c}\left(
s\right)  \label{a-6}
\end{multline}
and
\begin{multline}
\left( s+i\omega _{c}\right) A_{c}\left( s\right) +\mu _{c}I\left( s\right)
\sum_{c^{\prime }}\mu _{c^{\prime }}^{\ast }A_{c^{\prime }}\left( s\right) \\
=-\mu _{a}^{\ast }\mu _{c}J\left( s,\vec{r}\right) A_{0}\left( s\right) ,
\label{b-3}
\end{multline}
where
\begin{multline}
I\left( s\right) =\frac{1}{\left( 2\pi \right) ^{2}\hbar }\int
d^{3}k\,\sum_{\lambda } \\
\left( \widehat{\boldsymbol{p}}_{a}\cdot\widehat{\boldsymbol{\epsilon }}_{%
\vec{k},\lambda }\right) \left( \widehat{\boldsymbol{p}}_{a}\cdot\widehat{%
\boldsymbol{\epsilon }}_{\vec{k},\lambda }\right) \frac{\omega_{k}}{\left(
s+i\omega _{k}\right) },  \label{ia-1}
\end{multline}
and
\begin{multline}
J\left( s,\vec{r}\right) =\frac{1}{\left( 2\pi \right) ^{2}\hbar }\int
d^{3}k\sum_{\lambda } \\
\left( \widehat{\boldsymbol{p}}_{d}\cdot\widehat{\boldsymbol{\epsilon }}_{%
\vec{k},\lambda }\right) \left( \widehat{\boldsymbol{p}}_{a}\cdot\widehat{%
\boldsymbol{\epsilon }}_{\vec{k},\lambda }\right) \frac{\omega _{k}e^{-i\vec{%
k}\cdot\vec{r}}}{\left( s+i\omega _{k}\right) }.  \label{ja-1}
\end{multline}
We now eliminate $A_{c}\left( s\right) $ from Eq.(\ref{b-3}) and  Eq.(\ref%
{a-6}), introduce the sum over the ionization states,
\begin{equation}
L\left( s\right) =\sum_{c}\frac{\left| \mu _{c}\right| ^{2}}{\left(
s+i\omega _{c}\right) }  \label{ns-1}
\end{equation}
and obtain
\begin{equation}
A_{0}\left( s\right) =\frac{1}{s+i\omega _{0}+\left| \mu _{a}\right|
^{2}I\left( s\right) U\left( s,\vec{r}\right) },  \label{a-7}
\end{equation}
where
\begin{equation}
U\left( s,\vec{r}\right) =1-\frac{L\left( s\right) J^{2}\left( s,\vec{r}%
\right) }{I\left( s\right) \left[ 1+L\left( s\right) I\left( s\right) \right]
}.  \label{u-1}
\end{equation}
Remember that the probability of survival of the atom in its excited state
is given by
\begin{equation}
P\left( t\right) =\left| A_{0}\left( t\right) \right| ^{2},  \label{pt-1}
\end{equation}
where
\begin{equation}
A_{0}\left( t\right) =\int_{\mathcal{C}}\frac{ds}{2\pi i}\,e^{st}A_{0}\left(
s\right) ,  \label{at-5}
\end{equation}
and $\mathcal{C}$ is the standard contour of integration for the inverse
Laplace transform.

We perform sum over polarizations and the angular part of the integrations
of Eq.(\ref{ia-1}), and Eq.(\ref{ja-1}):
\begin{equation}
I\left( s\right) =\frac{2}{3\pi \hbar c^{3}}\int_{0}^{\infty }d\omega
^{\prime }\frac{\omega ^{\prime ^{3}}}{\left( s+i\omega ^{\prime }\right) }.
\label{ia-5}
\end{equation}%
and
\begin{equation}
J\left( s,\vec{r}\right) =\frac{1}{2\pi \hbar c^{3}}\int_{0}^{\infty
}d\omega ^{\prime }\frac{\omega ^{\prime 3}}{\left( s+i\omega ^{\prime
}\right) }D\left( \omega ^{\prime }r/c\right) ,  \label{ja-5}
\end{equation}%
where
\begin{multline}
D\left( z\right) =\widehat{\boldsymbol{p}}_{d}\cdot \widehat{\boldsymbol{p}}%
_{a}\left( S\left( z\right) +T\left( z\right) \right)   \notag \\
{}+\left( \widehat{\boldsymbol{r}}\cdot \widehat{\boldsymbol{p}}_{d}\right)
\left( \widehat{\boldsymbol{r}}\cdot \widehat{\boldsymbol{p}}_{a}\right)
\left( S\left( z\right) -3T\left( z\right) \right) ,  \label{dz-1}
\end{multline}%
with
\begin{equation}
\begin{split}
S\left( z\right) & =\frac{1}{2}\int_{-1}^{1}d\xi e^{-iz\xi }=\frac{\sin z}{z}%
, \\
T\left( z\right) & =\int_{-1}^{1}d\xi e^{-iz\xi }\xi ^{2}=-S^{\prime \prime
}\left( z\right) .
\end{split}
\label{dz-2}
\end{equation}%
Also we replace in Eq.(\ref{ns-1}), $\sum_{c}\rightarrow \int_{\omega
_{I}}^{\infty }d\omega _{c}\rho \left( \omega _{c}\right) $, where $\rho
\left( \omega _{c}\right) $ is the density of states for ionization of the
detector's atom, and write
\begin{equation}
L\left( s\right) =\int_{\omega _{I}}^{\infty }d\omega ^{\prime }\rho \left(
\omega ^{\prime }\right) \frac{\left\vert \mu _{c}\right\vert ^{2}}{\left(
s+i\omega ^{\prime }\right) }.  \label{ns-5}
\end{equation}%
We can now estimate the orders of magnitude of the functions in Eq.(\ref{u-1}%
). By inspection we have $I\sim J\sim \omega _{0}^{3}/\hbar c^{3}$, and $%
L\sim \left\vert \mu _{c}\right\vert ^{2}\rho \left( \omega \right) \sim
e^{2}a^{2}/\omega _{0}$, where $e$ is the electron's charge, and $a$ is of
the order of Bohr radius. Further since $e^{2}/a\sim \hbar \omega _{0}$, and
$c/\omega _{0}\sim 2\pi /\lambda $, the radiation wavelength, $LI\sim \left(
e^{2}/a\right) /\hbar \omega _{0}\left( a\omega _{0}/c\right) ^{3}\sim
\left( 2\pi a/\lambda \right) ^{3}\ll 1$, and we can neglect $LI$ with
respect to one in the denominator of Eq.(\ref{u-1}). Also we notice that $%
\left\vert \mu _{a}\right\vert ^{2}I/\omega _{0}\sim \left( 2\pi a/\lambda
\right) ^{3}\ll 1$.

It is instructive at this point to investigate Eq.(\ref{at-5}) in the \emph{%
absence} of the detector. We write explicitly
\begin{equation}
A_{0}^{\left( 0\right) }\left( t\right) =\int_{c}\frac{ds}{2\pi i}e^{st}%
\frac{1}{s+i\omega _{0}+\left| \mu _{a}\right| ^{2}I\left( s\right) },
\label{at-6}
\end{equation}
where $I\left( s\right) $ is given by Eq.(\ref{ia-5}). This is the familiar
Weisskopf--Wigner expression for the survival amplitude of an excited atom
in vacuum radiation field. Except for very short times, which is the main
interest of previous studies of the quantum Zeno effect, and very long times%
\cite{Khalfin1958}, the Weisskopf--Wigner approximation for $A_{0}\left(
t\right) $ is valid: one can safely substitute $s=-i\omega _{0}+\gamma $,
with $\gamma \ll \omega _{0}$, in Eq.(\ref{ia-5}), and evaluate the integral
of Eq.(\ref{at-6}). Discarding the imaginary part of the integral in Eq.(\ref%
{ia-5}), which just contributes an energy shift, we obtain
\begin{equation}
I\left( s\right) =\frac{2\omega _{0}^{3}}{3\hbar c^{3}},  \label{ia-7}
\end{equation}
which yields
\begin{equation}
A_{0}^{\left( 0\right) }\left( t\right) =e^{-\Gamma t/2-i\omega _{0}t},
\label{at-7}
\end{equation}
and the survival probability
\begin{equation}
P^{\left( 0\right) }\left( t\right) =e^{-\Gamma t}.  \label{pt-2}
\end{equation}
Here $\Gamma =4\omega _{0}^{3}\left| \mu _{a}\right| ^{2}/3\hbar c^{3}$ is
the Weisskopf--Wigner relaxation rate, or Einstein $A$ coefficient, for the
excited state of the atom.

We return now to Eq.(\ref{at-5}), reinstate the contribution of the
detector, and write
\begin{multline}
A_{0}\left( t\right) =  \notag \\
\int_{C}\frac{ds}{2\pi i}\,e^{st}\frac{1}{s+i\omega _{0}+\left\vert \mu
_{a}\right\vert ^{2}I\left( s\right) U\left( s,\vec{r}\right) }.  \label{k}
\end{multline}%
In the spirit of the Weisskopf--Wigner procedure we substitute $s=-i\omega
_{0}+\gamma $, with $\gamma \ll \omega _{0}$, in Eq.(\ref{ja-5}), and Eq.(%
\ref{ns-5}), and obtain
\begin{equation}
J\left( s,\vec{r}\right) =\frac{\omega _{0}^{3}}{4\pi \hbar c^{3}}D\left(
\omega _{0}r/c\right) ,  \label{ja-6}
\end{equation}%
and
\begin{equation}
L\left( s\right) =\pi \left\vert \mu _{c}\right\vert ^{2}\rho \left( \omega
_{0}\right) .  \label{ns-3}
\end{equation}%
Again with Eq.(\ref{pt-1}) the probability of the survival of the excited
state is then
\begin{equation}
P\left( t\right) =e^{-\Gamma U(\vec{r})t},  \label{pt-3}
\end{equation}%
where
\begin{equation}
U(\vec{r})=1-\frac{9}{64\pi ^{2}}\beta D^{2}\left( \omega _{0}r/c\right) ,
\label{u-2}
\end{equation}%
is the \emph{reduction factor} due to a \emph{single} detector atom at
position $\vec{r}$, and
\begin{equation}
\beta =2\pi \omega _{0}^{3}\left\vert \mu _{c}\right\vert ^{2}\rho \left(
\omega _{0}\right) /3\hbar c^{3}\sim \left( 2\pi a/\lambda \right) ^{3}.
\label{beta-1}
\end{equation}%
When the detector is in the far field region, we obtain
\begin{equation}
U(\vec{r})=1-\frac{9\beta }{4}\frac{c^{2}}{\omega _{0}^{2}r^{2}}l^{2}\sin
^{2}\omega _{0}r/c,  \label{u-3}
\end{equation}%
where $l=\widehat{\boldsymbol{p}}_{d}\cdot \widehat{\boldsymbol{p}}%
_{a}-\left( \widehat{\boldsymbol{r}}\cdot \widehat{\boldsymbol{p}}%
_{d}\right) \left( \widehat{\boldsymbol{r}}\cdot \widehat{\boldsymbol{p}}%
_{a}\right) $ is the dipole-dipole geometric factor, while for a detector at
$r\rightarrow 0$, we have $U(\vec{r})=1-\beta \left( \widehat{\boldsymbol{p}}%
_{d}\cdot \widehat{\boldsymbol{p}}_{a}\right) ^{2}$.

We observe that the irreversibility of the decay process can be understood
as due the continuum of the ionization states of the detector's atom. This
could also explain why there would not be coherent reflection of the
excitation back to the radiated atom. This allows us to avoid including an
explicit irreversibility in the model, just as the continuum of photon
states enabled Weisskopf and Wigner to get away with a discussion based
entirely on wave functions.

Notice that since $\beta $ is very small, the reduction due to one atom in
the detector is quite small, however when the detector is comprised of many
atoms the reduction can be significant.

\section{Discussion}

Our Hamiltonian Eq.(\ref{Hd-2}) for a many atom detector allows coherent
transfer of ionization from one detector atom to another. In practice, we do
not expect such coherent transfer to occur because the electron produced by
ionization is removed in the course of measurement and replaced by some
other externally supplied electron. If we neglect the coherent transfer and
assume all the atoms of the detector to be identical, we may conjecture that
we can simply replace Eq.(\ref{u-3}) by
\begin{equation}
U=1-\frac{9}{4}\beta \sum_{i}\frac{c^{2}}{\omega _{0}^{2}r_{i}^{2}}%
l_{i}^{2}\sin ^{2}\omega _{0}r_{i}/c,  \label{u-4}
\end{equation}
where $l_{i}=\widehat{\boldsymbol{p}}_{d}\cdot\widehat{\boldsymbol{p}}%
_{a}-\left( \widehat{\boldsymbol{r}}_{i}\cdot\widehat{\boldsymbol{p}}%
_{d}\right) \left( \widehat{\boldsymbol{r}}_{i}\cdot \widehat{\boldsymbol{p}}%
_{a}\right) $, and $\vec{r}_{i}$ is the location of the $i-th$ atom.
Averaging over the directions in space and detector atom polarizations, we
find $\langle \overline{l^{2}}\rangle =2/7$. For a thin spherical shell
detector of radius $R$, and $N_{a}$ atoms, the reduction factor, Eq.(\ref%
{u-4}), is
\begin{equation}
U=1-\frac{9}{14}\beta N_{a}\frac{c^{2}}{\omega _{0}^{2}R^{2}}\sin ^{2}\omega
_{0}R/c.  \label{u-5}
\end{equation}
The efficiency of the detector is hidden within $\beta$. If $U \approx 1$,
the coupling between the detector atoms should not have been neglected. A
more detailed calculation would have to include the transfer of electrons
into and out of the detector array and would have to be described by an
equation of motion, containing non-Hamiltonian terms, for a density matrix.

In conclusion, we observe that to inhibit the decay of an unstable state,
the latter need be just constantly watched by appropriate detectors. No
active measurements, perturbing the atom, are needed, and neither is the
early time evolution of the atom crucial for decelerating (or accelerating)
the decay rate. Although it may not be true that \textquotedblleft A watched
pot never boils\textquotedblright \cite{Gaskell1848}, we have shown that at
least a watched pot boils more slowly.

\begin{acknowledgments}
One of us (A.R.) would like to acknowledge very helpful discussions with A.
Ben-Kish, O. Firstenberg, A. Fisher, and M. Shuker,
\end{acknowledgments}


\begin{thebibliography}{99}
\bibitem{Eschner2001} See recent paper by J. Eschner, Ch. Raab, F.
Schmidt-Kater, and R. Blatt, Nature \textbf{413}, 495, (2001), and
references cited therein.

\bibitem{Koshino2005} See recent review by K. Koshino, and A. Shimizu,
Physics Report \textbf{412}, 191, (2005), and references cited therein.

\bibitem{Purcell1946} E. M. Purcell, Phys. Rev. \textbf{69}, 681, (1946).

\bibitem{Kleppner1971} D. Kleppner, in \emph{Atomic Physics and Astrophysics}%
, edited by M. Chretien and E. Lipworth (Gordon and Breach, NY, 1971), Sec.
6.3, p. 5.

\bibitem{Kleppner1981} Daniel Kleppner, Phys. Rev. Lett. \textbf{47}, 233,
(1981).

\bibitem{Drexhage1974} K. H. Drexhage, in \emph{Progress in Optics}, edited
by E. Wolf (North-Holland, Amsterdam, 1974), Vol. 12, p. 165.

\bibitem{Hulet1985} R. G. Hulet, E.S. Hilfer, and D. Kleppner, Phys. Rev.
Lett. \textbf{55}, 2137, (1985).

\bibitem{Heinzen1987} D. J. Heinzen, J. J. Childs, J. E. Thomas, and M. S.
Feld, Phys. Rev. Lett. \textbf{58}, 1320, (1987).

\bibitem{Misra1977} B. Misra ans E.C.G. Sudarshan, J. Math. Phys. \textbf{18}%
, 756 (1977).

\bibitem{Itano1990} W.M. Itano, D.J. Heinzen, J.J. Bollinger, and D.J.
Wineland, Phys. Rev. A \textbf{41}, 2295 (1990).

\bibitem{Schenzle1991} See e.g., V. Frerichs and A. Schenzle, Phys Rev. A
\textbf{44}, 1962 (1991), and E. Block and P.R. Berman, Phys. Rev. A \textbf{%
44}, 1466 (1991), and M.J. Gagen and G.J. Milburn, Phys. Rev. A \textbf{47},
1467 (1992).

\bibitem{Fischer2001} M.C. Fischer, B. Guti\'{e}rrez-Medina, and M.G.
Raizen, Phys. Rev. Lett. \textbf{87}, 040402 (2001).

\bibitem{Khalfin1958} L.A. Khalfin, Zh. Eksp. Theo. Fiz. \textbf{33}, 1371
(1958) [Sov. Phys. JETP \textbf{6}, 1503 (1958)]

\bibitem{Peres1980b} A. Peres, Ann. Phys. NY \textbf{129}, 33,(1980).

\bibitem{Facchi2001} P. Facchi, H. Nakazato, and S. Pascazio, Phys. Rev.
Lett. \textbf{86}, 2699 (2001), and P. Facchi, and S. Pascazio, in \textit{%
Progress in Optics,} Ed. E. Wolf (Elsevier, Amsterdam, 2001) \textbf{42},
147.


\bibitem{WW1930} V. Weisskopf and E. Wigner, Z. Phys. \textbf{63}, 54 (1930).

\bibitem{Gaskell1848} Elizabeth C. Gaskell, \textit{Mary Barton, a tale of
Manchester life} xxxi (1848): ``What's the use of watching? A watched pot
never boils.''
\end{thebibliography}

\end{document}